\documentclass[aps,pra,longbibliography,superscriptaddress,twocolumn,twocolumn,10pt]{revtex4-2}
\usepackage{amsmath}
\usepackage{latexsym}
\usepackage{graphicx}
\usepackage{amsfonts}
\usepackage{braket}
\usepackage{hyperref}
\usepackage{natbib}
\usepackage{amssymb}
\usepackage{nicefrac}
\usepackage{fancyhdr}
\usepackage[utf8]{inputenc}
\usepackage{dsfont}
\usepackage{colortbl}
\usepackage{xcolor}
\usepackage{subfigure}
\usepackage{psfrag}
\usepackage{tikz}
\usepackage{paralist}
\usepackage{nicefrac}
\usepackage{ulem}
\usepackage{placeins}
\usepackage{mathtools}
\usepackage{bbm}
\usepackage{color}
\usepackage{siunitx}
\makeindex

\begin{document}

\title{Simulation and Benchmarking of Real Quantum Hardware}

\author{T. Piskor}
\email{tomislav.piskor@eviden.com}
\affiliation{science + computing AG / Eviden, Hagellocher Weg 73, 72070 Tübingen, Germany}
\author{M. Schöndorf}
\affiliation{science + computing AG / Eviden, Hagellocher Weg 73, 72070 Tübingen, Germany}
%\affiliation{BWI GmbH, I\&T TM\&I TC Quantum-Enabled Technologies, Lurgiallee 10, 60439 Frankfurt am Main}
\author{M. Bauer}
\author{D. Smith}
\affiliation{science + computing AG / Eviden, Hagellocher Weg 73, 72070 Tübingen, Germany}
\author{T. Ayral}
\affiliation{Eviden Quantum Lab, 78340 Les Clayes-sous-Bois, France}
\author{S. Pogorzalek}
\author{A. Auer}
\affiliation{IQM Quantum Computers, Georg-Brauchle-Ring 23-25, 80992 Munich, Germany}
\author{M. Papi\v{c}}
\affiliation{IQM Quantum Computers, Georg-Brauchle-Ring 23-25, 80992 Munich, Germany}
\affiliation{Department of Physics and Arnold Sommerfeld Center for Theoretical Physics,
Ludwig-Maximilians-Universität München, Theresienstr. 37, 80333 Munich, Germany}

\begin{abstract}
The effects of noise are one of the most important factors to consider when it comes to quantum computing in the noisy intermediate-scale quantum computing (NISQ) era that we are currently in. Therefore, it is important not only to gain more knowledge about the noise sources appearing in current quantum computing hardware in order to suppress and mitigate their contributions, but also to evaluate whether a given quantum algorithm can achieve reasonable results on a given hardware. To accomplish this, we need noise models that can describe the real hardware with sufficient accuracy. Here, we present a noise model that has been evaluated on superconducting hardware platforms and could be adapted to other common architectures such as trapped-ion or neutral atom devices. We then benchmark our model by simulating a 20-qubit superconducting quantum computer, and compare the accuracy of our model to similar approaches from the literature and demonstrate an improvement in the overall prediction accuracy.

\end{abstract}

\maketitle

\section{Introduction}
\label{sec:Introduction}

Quantum computing is a promising candidate to solve complex problems in various fields such as optimization \cite{PhysRevX.10.021067, farhi2014quantumapproximateoptimizationalgorithm, peruzzo2014variational, abbas2023quantumoptimizationpotentialchallenges}, quantum simulation \cite{RevModPhys.86.153, daley2022practical, fauseweh2024quantum}, etc. much more efficiently than classical computers. However, the actual realization of quantum devices with a sufficient number of qubits to exploit the true potential of quantum computing is proving to be a major challenge.

There are various hardware realizations that are currently used to implement quantum computing units, with the most established realizations currently being superconducting qubits \cite{clarke2008superconducting,brecht2016multilayer,you2011atomic}, ions \cite{PhysRevLett.74.4091, HAFFNER2008155, RevModPhys.75.281} and neutral atoms \cite{Henriet2020quantumcomputing, bluvstein2024logical}. There are several other strategies, ranging from semiconductor qubits \cite{chatterjee2021semiconductor, burkard2023semiconductor} to more exotic approaches such as Majorana qubits \cite{flensberg2021engineered, aguado2020majorana}, yet it is not clear which will be the best solution in the long run.

All current realizations of quantum computers are still considered noisy intermediate-scale quantum (NISQ) devices \cite{cheng2023noisy, bharti2022noisy, preskill2018quantum}, meaning that the hardware is noisy and can only be used for a limited number of error-prone operations before the quantum nature of the system vanishes \cite{bharti2022noisy}. Therefore, understanding, mitigating and suppressing the noise is one of the main focuses in the improvement of current quantum computers.

Due to the limited number of operations, it is not only necessary to evaluate the optimal hardware realization to tackle a specific problem on a quantum computer, but there is also a need to understand and characterize the noise \cite{breuer2002theory, papivc2023fast} in the devices in order to mitigate its effects \cite{cai2023quantum, kim2023evidence}. For both problems, good noise models that accurately describe the quantum device are crucial.

% In order to overcome this hurdle the need for a better understanding of the noise has spurred the development of novel characterization techniques \cite{breuer2002theory, papivc2023fast}, as well as noise suppression and mitigation approaches\cite{cai2023quantum, kim2023evidence}. Especially, for the latter a reliable noise model is required for several approaches.

We validate our model by comparing the predictions to the results obtained from an IQM superconducting 20-qubit chip~\cite{ronkko2024premises} and to the results already performed on IBM Q Melbourne~\cite{Georgopoulos2021}. The noise model described in this paper was implemented using the Eviden Qaptiva software framework, and the simulation was subsequently run on the Eviden Qaptiva system \cite{Qaptiva}. For the comparison itself, we used specific benchmark circuits varying in type (structured and random) and size (from 2 to 15 qubits and circuit depths ranging from 5 to 1062). This means that our model permits us to predict measurement results for arbitrary quantum algorithms running on real hardware. In both cases, we show that the results of the simulations are in good agreement with the results from real hardware. \\
\hspace*{1em}We construct a general and holistic noise model that allows to describe noisy quantum computers with an arbitrary number of qubits and gates. Moreover, even though our model has been evaluated on superconducting devices, it can be adapted to different platforms by using different parameters, pertinent to that hardware. Adaptation to different hardware realizations can be easily achieved by feeding in the noise parameters of the respective hardware. The model itself is a combination of different error channels that are most likely to appear in all hardware realizations of quantum computers. \\
\hspace*{1em}The paper is organized as follows: In Sec.~\ref{sec:iqm_qpu} we describe the IQM 20-qubit chip used for the benchmarks. The associated noise model is discussed in Sec.~\ref{sec:noise_model} and the results of simulating the noise model against the real hardware are shown in Sec.~\ref{sec:benchmarks}. In Sec.~\ref{sec:conclusion} we present the conclusion of our work.

\section{The IQM 20-Qubit Chip}
\label{sec:iqm_qpu}
\begin{figure}[]
\includegraphics[width=.45\textwidth]{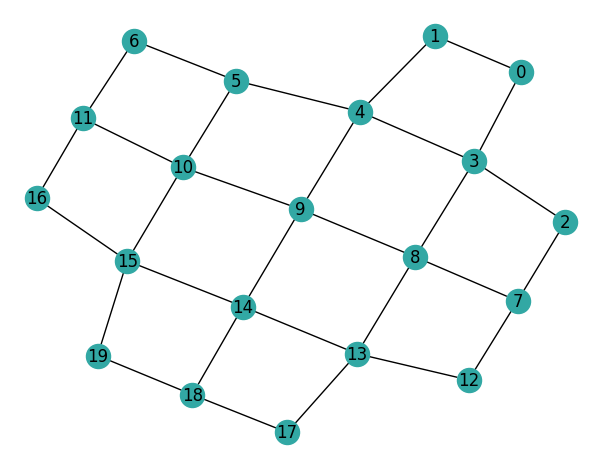}
\caption{Setup of the IQM chip used to benchmark the simulated results. Every node denotes a superconducting Transmon qubit. Tunable couplers between qubits are denoted by lines. The qubits are arranged in a square grid.}
\label{fig:IQM topology}
\end{figure}
In this paper, the measurement results used to validate the simulation results were obtained from specific hardware. In particular, a superconducting 20-qubit chip fabricated by IQM was used to run the benchmark circuits presented in Sec.~\ref{sec:benchmarks}.

Regarding the topology, the qubits are arranged in a square grid topology (see Fig.~\ref{fig:IQM topology}). Whenever a two-qubit gate is applied between non-neighboring qubits, a topology adaptation step (e.g., SWAP insertion \cite{saeedi2011synthesis}) is required. To reduce crosstalk between the qubits, tunable couplers are placed between all connected qubits~\cite{PRXQuantum.4.010314}. These tunable couplers are also used to implement the native entangling gate - the \texttt{CZ} gate, described by the unitary:

\begin{align}
\texttt{CZ} = 
    \begin{pmatrix}
        1 & 0 & 0 & 0 \\
        0 & 1 & 0 & 0 \\
        0 & 0 & 1 & 0 \\
        0 & 0 & 0 & -1 \\
    \end{pmatrix}.
\end{align}

The single-qubit gate implemented on the hardware describes an arbitrary rotation of angle $\Theta$ around the $\cos(\Phi)\hat\sigma_x + \sin(\Phi)\hat\sigma_y$ axis on the Bloch sphere \cite{krantz2019quantum, McKay_2017}. The terms $\hat\sigma_x$ and $\hat\sigma_y$ represent the Pauli X and Y matrices defined as
\begin{align}
\label{eq:paulix}
\hat \sigma_x &= \begin{pmatrix}
    0 & 1\\
    1 & 0
\end{pmatrix}, \\
\label{eq:pauliy}
\hat \sigma_y &= \begin{pmatrix}
    0 & -i\\
    i & 0
\end{pmatrix},
\end{align}
%as defined in~\ref{eq:paulix} and~\ref{eq:pauliy}, respectively
and $\Phi$ represents the azimuthal angle on the Bloch sphere.
Here the gate is denoted as $\texttt{PRX}(\Theta, \Phi)$ and its matrix representation is given by:
\begin{align}
\texttt{PRX}(\Theta,\Phi) = 
    \begin{pmatrix}
        \cos\frac{\Theta}{2} & -i\rm{e}^{-i\Phi}\sin\frac{\Theta}{2} \\
         -i\rm{e}^{i\Phi}\sin\frac{\Theta}{2} & \cos\frac{\Theta}{2}
    \end{pmatrix}.
\end{align}
This gate is sometimes named differently in the literature, e.g. $\texttt{R}$ in Qiskit~\cite{qiskit} and \texttt{PhasedX} in pytket~\cite{pytket}. The two gates \texttt{PRX} and \texttt{CZ} form a complete gate set for the IQM quantum computing hardware~\cite{kaye2006introduction}. 

Like any quantum computing device, the IQM 20-qubit chip~\cite{papivc2023fast} is affected by noise. Noise is characterized by different noise parameters that must be determined for the respective hardware~\cite{nielsen2010quantum}. The characteristic quantities measured for the noise model, which have been obtained by performing coherence time measurements and randomized benchmarking experiments~\cite{iqm20qpu}, are listed below:
\begin{itemize}
    \item Single-qubit gate fidelity $\mathcal{F}_\texttt{PRX}$ and time $T_\texttt{PRX}$
    \item Two-qubit gate fidelity $\mathcal{F}_\texttt{CZ}$ and time $T_\texttt{CZ}$
    \item Relaxation time $T_1$
    \item Depolarization time $T_2$
    \item Measurement error probabilities for state 0 $\epsilon_{\rm meas}^{0}$ and state 1 $\epsilon_{\rm meas}^{1}$
\end{itemize}
These parameters together with the corresponding gate durations are used as inputs for the  the noise model, which implements a digital twin of the IQM 20-qubit chip. A detailed description of this procedure is given in Sec.~\ref{sec:noise_model}. In particular, all input parameters were determined for each qubit or qubit pair in order to obtain a more accurate simulation of the hardware. The average parameters used for the IQM 20-qubit chip are listed in Table~\ref{tab:noise_parameters}. The parameters for all qubits and qubit pairs are given in App.~\ref{app2}.

\begin{figure}[]
\includegraphics[width=.45\textwidth]{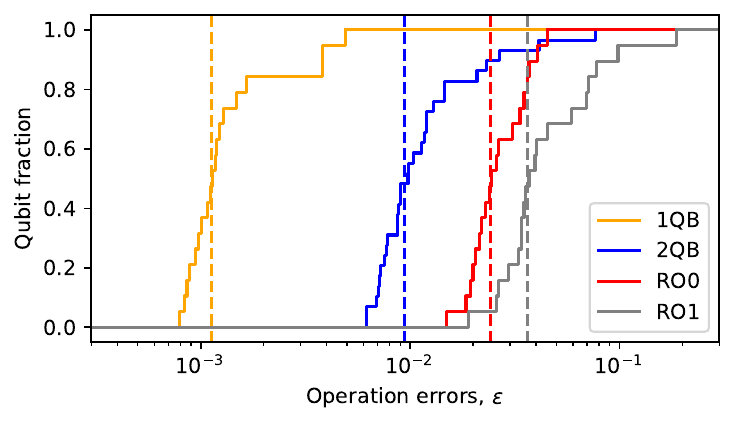}
\caption{Cumulative distribution of average single-qubit gate errors (1QB), two-qubit gate errors (2QB), readout errors of state 0 (RO0), and readout errors of state 1 (RO1) for the IQM 20-qubit QPU with vertical lines indicating median values.}
\label{fig:cdf}
\end{figure}

To gain further insight into the distribution of single- and two-qubit gate errors, as well as the readout errors, we consider the cumulative distributions of the single parameters, shown in Fig.~\ref{fig:cdf}. The details about the parameters obtained are given in~\cite{iqm20qpu}.

\begin{table}[]
    \vspace{0.5cm}
    \centering
    \begin{tabular}{c|c|c} \hline
        Parameter & Mean & Median \\  \hline
        $\mathcal{F}_\texttt{1QB}$ & \SI{99.85}{\%} & \SI{99.89}{\%}\\
        $\mathcal{F}_\texttt{2QB}$ & \SI{98.59}{\%} & \SI{99.06}{\%} \\
        $T_\texttt{1QB}$ & \SI{20}{ns} & \SI{20}{ns} \\
        $T_\texttt{2QB}$ & \SI{40}{ns} & \SI{40}{ns} \\
        $T_1$ & \SI{41.8}{\mu s} & \SI{43.1}{\mu s} \\
        $T_2$ & \SI{3.2}{\mu s} & \SI{2.8}{\mu s} \\
        $\epsilon_{\rm meas}^{0}$ & \SI{2.66}{\%} & \SI{2.43}{\%}\\
        $\epsilon_{\rm meas}^{1}$ & \SI{5.09}{\%} & \SI{3.63}{\%}\\ 
        \hline
    \end{tabular}
    \caption{Average noise parameters of the IQM 20-qubit chip, as measured during the calibration of the chip. Here, \texttt{1QB} denotes the single-qubit gate \texttt{PRX} and \texttt{2QB} denotes the two-qubit gate \texttt{CZ}. The average is taken over all connected qubit pairs for the two qubit gate fidelities and times. More details about the noise parameters can be found in App.~\ref{app2}.}
    \label{tab:noise_parameters}
\end{table}

\section{The noise model}
\label{sec:noise_model}
Quantum hardware is affected by noise due to interactions with the environment, miscalibration, errors in the control electronics, etc. For techniques such as error mitigation, it is important to have an accurate model of the corresponding errors in the quantum hardware. Furthermore, a good hardware simulation model is essential for benchmarking current hardware and for evaluating whether certain algorithms run with a reasonable quality on the device. Here we describe a noise model that was originally intended to model the IQM superconducting hardware described in Sec.~\ref{sec:iqm_qpu}, but can be adapted to be used on any other common quantum hardware implementation (e.g. ions, neutral atoms, etc.) by changing the respective noise parameters.

In order to balance the accuracy and generality of the noise model, we focus on noise channels (see next section) that typically appear in most current quantum hardware. In the following subsections, we present and discuss the noise channels used in our model description.

Each quantum channel $\mathcal{E}$ described in Secs.~\ref{subsec:relaxation}-\ref{subsec:measurement errors} imposes a transformation of the density matrix $\hat\rho$ representing the quantum system (in this case, the state of the quantum register \cite{nielsen2010quantum}),
\begin{align}
\hat\rho \longrightarrow \mathcal{E}(\hat \rho).
\end{align}
The quantum map $\mathcal{E}$ has the properties of being linear, completely positive (CP), and trace-preserving (TP). These ensure that the properties of the density matrix which are necessary to describe valid quantum states are preserved.

We use the Kraus operator formalism~\cite{kraus1983operations} to describe the channel $\mathcal{E}$, i.e.

\begin{equation}
    \hat\rho \rightarrow \hat\rho^\prime = \sum_i \hat K_i \hat\rho \hat K_i^\dagger,
\end{equation}

where $K_i$ corresponds to the Kraus operators. In the next subsections, three different channels that are used within our noise model with their corresponding Kraus operators will be presented. 

\subsection{Energy Relaxation}
\label{subsec:relaxation}
Since the qubits of a quantum hardware are not perfectly isolated, there is always some residual interaction with the environment. One of the processes resulting from these interactions is referred to as amplitude damping \cite{carroll2022dynamics, breuer2002theory}. In this process, the energy of the qubit is transferred to its environment, leading to a decay of the qubit from its excited to its ground state. The typical characteristic quantity of these processes is called the $T_1$ time which denotes a characteristic time scale for a qubit to relax to the ground state under the influence of the environment. In our implementation, $T_1$ is also the input parameter of the model. %To describe the process quantitatively, we need the quantum channel in the \emph{Kraus representation} \cite{kraus1983operations}, also called \emph{operator-sum representation} \cite{nielsen2010quantum}. The Kraus operators applied to the density matrix of the system of interest characterize the time evolution of the system under the influence of the respective channel. 
A relaxation process with a characteristic timescale of $T_1$, is described by the following Kraus operators
\begin{align}
\label{eq:kraus_relaxation}
\hat K_{r,0} &= \ket{0}\bra{0} + \sqrt{1-p_r(t)} \ket{1}\bra{1}  \\
\hat K_{r,1} &= \sqrt{p_r(t)}\ket{0}\bra{1},
\end{align}
with $p_r(t) = 1-\exp(-t/T_1)$ as the time-dependent relaxation rate and $t$, in the context of noise modeling, is determined by the duration of the operation whose noise we are describing with this process.

\subsection{Dephasing}
\label{subsec:dephasing}
Another significant environmental factor that impacts qubits is dephasing. In this scenario, the qubit does not actively exchange energy with its environment, as observed in the relaxation case, but rather undergoes a decay in its phase information over time~\cite{chen2019quantifying, breuer2002theory}. This phenomenon can be attributed to the presence of a frequency background of the environment, resulting in time-dependent fluctuation of the qubit's characteristic frequency. This process leads to a damping of the off-diagonal elements of the density matrix, i.e. dephasing. It is important to note that relaxation also results in the  dephasing of the qubit since relaxation to the ground state effectively destroys phase information as well. The characteristic quantity that describes the total dephasing is referred to as the $T_2$ time, which is defined as the average time it takes for a qubit to fully lose its phase information. This $T_2$ encompasses both the dephasing arising from relaxation processes and the pure dephasing component. The pure dephasing component is characterized by its own parameter, $T_\phi$. The overall $T_2$ time combines the dephasing due to $T_1$ processes and the pure dephasing, and is given by $1/T_2 = 1/(2T_1) + 1/T_{\phi}$ \cite{krantz2019quantum}. Since the $T_1$ component of dephasing is already incorporated into the relaxation, we proceed to define the Kraus operators for the pure dephasing process, which is expressed as follows:
\begin{align}
\label{eq:kraus_dephasing}
\hat K_{d,0} &= \sqrt{1-p_d(t)}  \mathds{\hat 1} \\
\hat K_{d,1} &= \sqrt{p_d(t)} \hat \sigma_z,
\end{align}
where $p_d(t) = 1-\exp(-t/T_2)$, and $\hat \sigma_z$ is the Pauli Z matrix.
%\mariuscom{TODO: Add appendix for 1/f noise and comment here}

\subsection{Depolarization}
\begin{figure*}[t!]
\includegraphics[width=.9\textwidth]{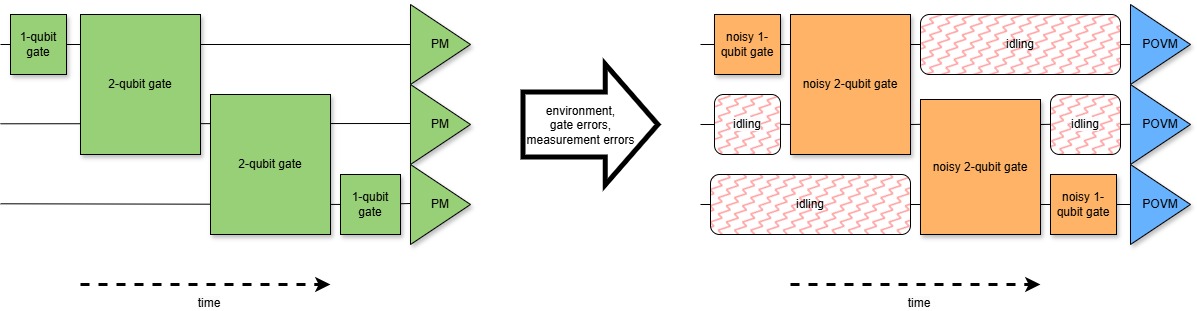}
\caption{Transition from an ideal to a noisy circuit used in our simulations. Each noiseless gate (solid boxes) is transferred to a quantum channel that includes the effect of the gate and the depolarization error. When the qubit is in the idle state (zigzag boxes), the relaxation and dephasing channel is applied and the projective measurements (triangles) of the noiseless circuit are replaced by a more general POVM accounting for the measurement errors.}
\label{fig:noise model}
\end{figure*}

\label{subsec:depolarization}
% In order to account for any remaning errors, we use the quantum depolarizing channel.
The quantum error channel employed in our model to describe gate errors is the so-called depolarization channel. %This is regarded as the most general error that can occur on a qubit, i.e., transferring a specific qubit state to a completely mixed state~\cite{nielsen2010quantum}. Given the complexity of most gate operations in real hardware and the difficulty in characterizing the exact error channel acting on the qubit, it is a valid approach to assume that the error induced during the application of the gate is of a depolarization kind. In this approach, the depolarization channel is constructed such that the parameter fed into the model is the average process fidelity, which is calculated as the gate fidelity measured for the quantum computing hardware. The Kraus operators are defined as follows:
A full tomographic characterization of the quantum channel of a gate is resource intensive~\cite{tomography}. Unlike randomized benchmarking, which is routinely performed as part of the calibration procedure, tomography is not routinely performed and therefore not necessarily available to the user. For lack of such tomographic information, we therefore choose to describe the errors of quantum gates with the quantum depolarizing channel that does not assume any predominance of an error (X, Y or Z) over another. It is defined as:
\begin{align}
\label{eq:kraus_dephasing}
\hat K_{{\rm dep}, 0} &= \sqrt{1-p_{\rm dep}(t)} \mathds{\hat 1}, \\
\hat K_{{\rm dep}, 1} &= \sqrt{p_{\rm dep}(t)/3} \hat \sigma_x,\\
\hat K_{{\rm dep}, 2} &= \sqrt{p_{\rm dep}(t)/3} \hat \sigma_y,\\
\hat K_{{\rm dep}, 3} &= \sqrt{p_{\rm dep}(t)/3} \hat \sigma_z,
\end{align}
with $\mathds{\hat 1}$ being the identity operator, the Pauli operators $\hat\sigma_x$ and $\hat\sigma_y$ defined in~\eqref{eq:paulix} and~\eqref{eq:pauliy} and the Pauli Z operator $\hat\sigma_z$ defined as
\begin{align}
\label{eq:pauliz}
\hat \sigma_z &= \begin{pmatrix}
    1 & 0\\
    0 & -1
\end{pmatrix}.
\end{align}
We fix the depolarization probability to match the average error rates we introduced above.
We also note that in the case of random circuits, the use of depolarizing noise instead of a more specific noise model is justified by the fact that averaging over random gates turns any noise channel into a depolarizing channel with equivalent average process error.
%\mariuscom{TODO: give formula for $p_{\rm dep}$}

\subsection{State Preparation and Measurement (SPAM) Errors}
\label{subsec:measurement errors}
The last error we include in our modeling of noisy circuits is the error in the final measurement of the qubits. In particular, registering an erroneous result means falsely measuring the qubit in the ground state when it is actually in the excited state, or the opposite case of falsely measuring the excited state. Imperfect measurements in quantum devices are a very common source of error in current quantum architectures (see Fig. \ref{fig:cdf}) and should be considered in a realistic modeling of quantum computing hardware. 
To model imperfect measurements, we replace the perfect projective quantum measurements by imperfect positive operator-valued measures (POVMs) in our model of the circuit \cite{nielsen2010quantum}. The measurement on each qubit is thus described by a pair of positive semi-definite operators $\hat E$ and $\textbf{1}-\hat E$ with respective outcomes $0$ and $1$. The probability of measuring the wrong result for a qubit in state $\ket{0}$ or $\ket{1}$ is denoted by $\epsilon_{\rm meas}^{0}$ or $\epsilon_{\rm meas}^{1}$, respectively.
Within our noise model, we neglected initialization or state preparation errors because these errors are generally difficult to track and -- at least during our studies -- did not show any major differences in our results when included. Consequently, we decided not to include this source of error and to consider only measurement errors for our noise model.

%\mariuscom{TODO: Add paragraph on initialization errors or call this SPAM errors and argue why not included}

\subsection{Modeling Noisy Circuits}
\label{subsec:modelling noisy circuits}
In the last subsection, we presented the building blocks for our noise model. To actually model a circuit executed on real noisy quantum computing hardware, we need to assemble these building blocks together.

The time evolution of the system is determined by applying the respective noise channels. The specific choice in which to apply them is the main characteristic of the respective noise model. The construction of the noise model is as follows:
\begin{itemize}
\item For each gate, the depolarizing channel is applied for the gate time $t_g$, where the error is characterized by the gate fidelity obtained from the hardware (measured via randomized benchmarking \cite{dankert2009exact, magesan2012characterizing, chow2009randomized, gaebler2012randomized}).
\item During the idle time of the qubit, i.e. when no gate is applied to the qubit, the relaxation and dephasing channels are applied to the qubit, characterized by $T_1$ and $T_2$, which are obtained from the hardware.
\item At the very end of the circuit we apply the measurement channel including measurement errors, where again the measurement error parameters $\epsilon_{\rm meas}^{0/1}$ are measured in the hardware.
\end{itemize}
A schematic description of our noise model, i.e. how a noiseless circuit is transformed into a noisy circuit, can be found in Fig.~\ref{fig:noise model}.

Georgopoulos et al.~\cite{Georgopoulos2021} present a comparable approach to simulate superconducting hardware by IBM. The main difference between our model and theirs is that in their model after a gate, the depolarizing channel is applied first and afterwards the relaxation and dephasing channels are applied to simulate idle noise. We apply these two channels during the idle time of the qubit, meaning the depolarizing channel is only applied for gate errors while relaxation and dephasing occurs only on idle qubits. From our point of view, this seems to be a physically more reasonable approach. The results presented in the next section also provide strong evidence for this assumption. The described noise model and all simulations presented in the next section were run on the Eviden Qaptiva 1.11.2, a quantum programming and simulation platform \cite{Qaptiva}.

\section{Noise Model Comparison to Real Hardware}
\label{sec:benchmarks}

\iffalse
\begin{figure*}
\includegraphics[width=.49\textwidth]{GHZ_2_measured_2025-6-12_calibrated_2025-6-12.pdf}
\includegraphics[width=.49\textwidth]{GHZ_3_measured_2025-6-12_calibrated_2025-6-12.pdf}
%\includegraphics[width=.49\textwidth]{GHZ_4_measured_2025-6-12_calibrated_2025-6-12.pdf}
%\includegraphics[width=.49\textwidth]{GHZ_5_measured_2025-6-12_calibrated_2025-6-12.pdf}
%\includegraphics[width=.49\textwidth]{GHZ_6_measured_2025-6-12_calibrated_2025-6-12.pdf}
%\includegraphics[width=.49\textwidth]{GHZ_7_measured_2025-6-12_calibrated_2025-6-12.pdf}
\caption{Comparison of the results of our noise model simulations and the measurement result obtained on the IQM 20-qubit QPU for the GHZ circuits: GHZ-2 (top left), GHZ-3 (top right), GHZ-4 (center left), GHZ-5 (center right), GHZ-6 (bottom left), and GHZ-7 (bottom right). To get the simulated results, a deterministic simulation was performed which generates a probability distribution for all states. The state probabilities have then been multiplied by 10,000 to get the counts.
The results on real hardware were obtained by repeating 10,000 circuit executions (shots) 50 times. The blue bars show the counts obtained from the simulation for the respective states, and the orange bars show the counts measured in the experiment. The error bars represent the standard deviations occurring from the repetitions of the measurements. States with counts below 100 have been neglected for visualization reasons.}
\label{fig:results_GHZ}
\end{figure*}
\fi

In this section we compare our noise model with measurement results obtained on a real IQM 20-qubit hardware (see Sec.~\ref{sec:iqm_qpu}). For the comparison, we choose several benchmark circuits varying from very small (5) to rather large ($>100$) circuit depths. 

\begin{figure*}
\includegraphics[width=.49\textwidth]{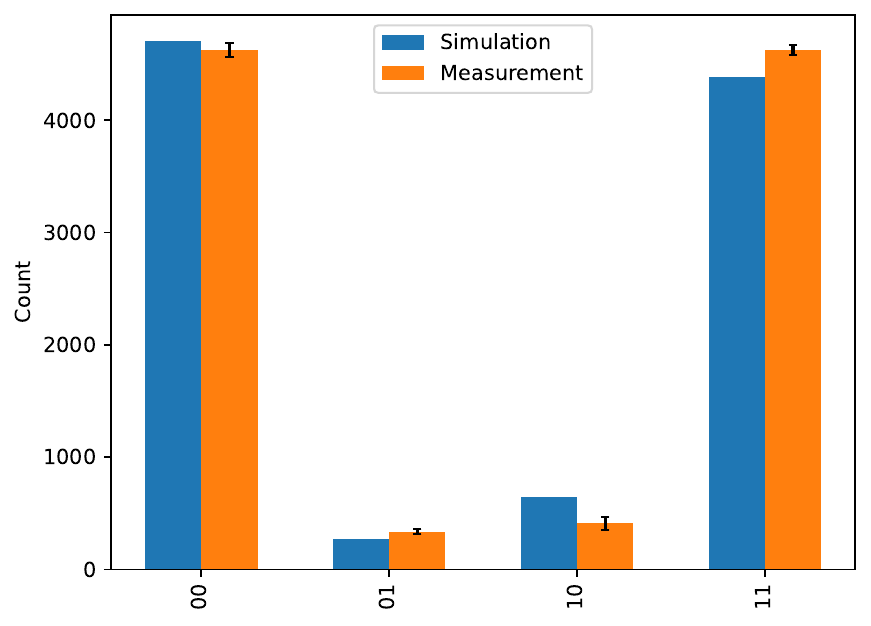}
\includegraphics[width=.49\textwidth]{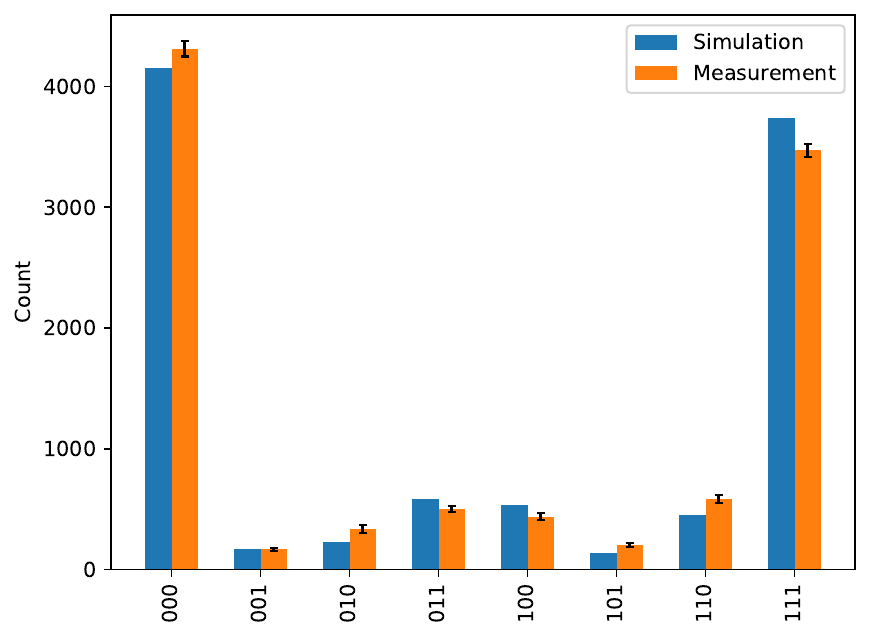}
\includegraphics[width=.49\textwidth]{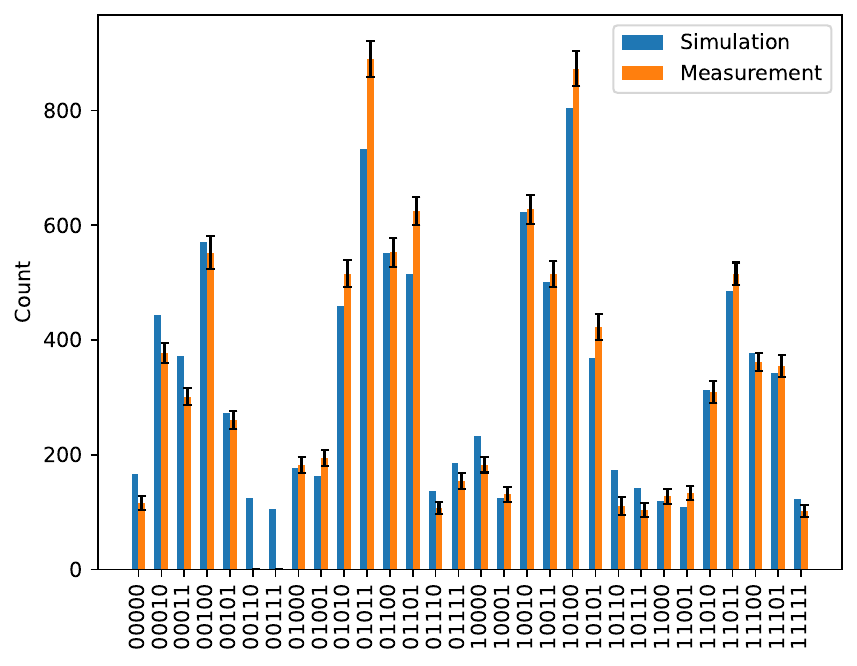}
\includegraphics[width=.49\textwidth]{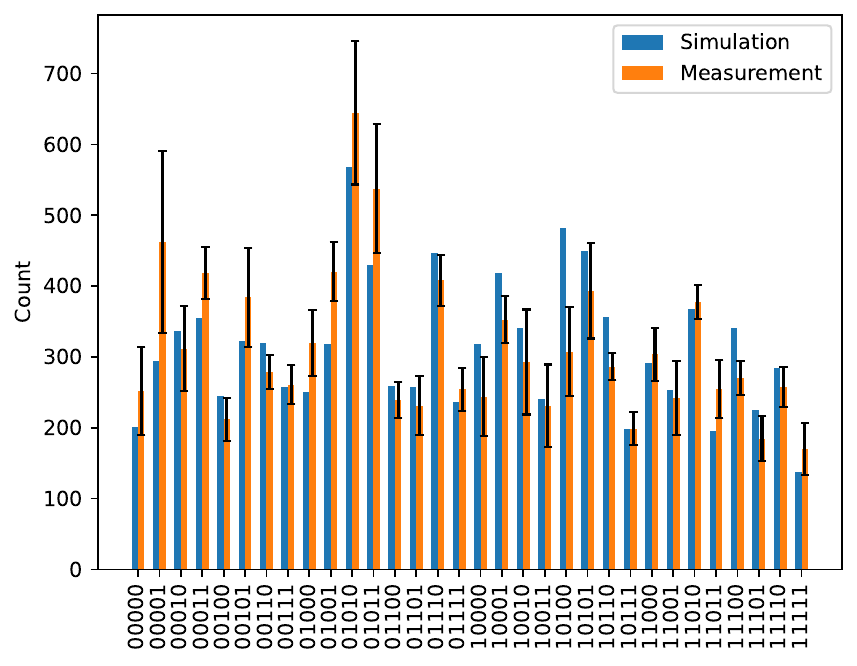}
\caption{Comparison of the results of our noise model simulations and the measurement results obtained on the IQM 20-qubit QPU for different classes of circuits: GHZ-2 (top left), GHZ-3 (top right), RU (bottom left), QAOA (bottom right). To get the simulated results, a deterministic simulation was performed which generates a probability distribution for all states. The state probabilities have then been multiplied by 10,000 to get the counts.
The results on real hardware were obtained by repeating 10,000 circuit executions (shots) 50 times. The blue bars show the counts obtained from the simulation for the respective states, and the orange bars show the counts measured in the experiment. The error bars represent the standard deviations due to finite sampling. States with counts below 100 have been neglected for visualization reasons.}
\label{fig:results_others}
\end{figure*}

\subsection{Comparison Metric: The Hellinger Distance}
\label{subsec:metric}

For a quantitative comparison, we need reasonable metrics to compare the probability distribution measured in the experiment with the one resulting from the simulation of our noise model. A widely used mathematical metric to compare the similarity of probability distributions is the Hellinger distance \cite{PhysRevA.97.062342}. The Hellinger distance for two probability distributions $P=\{p_i\}$ and $Q=\{q_i\}$ can be computed as follows:
\begin{align}
h(P,Q) = \frac{1}{\sqrt{2}} \sqrt{\sum_i^n \left(\sqrt{p_i} - \sqrt{q_i}\right)^2},
\end{align}
where $n$ is the number of discrete categories (bitstring outcomes) in the probability distribution.
The Hellinger distance assumes values in the range of $[0,1]$. A Hellinger distance of zero represents a perfect overlap between two distributions. Another important property of the Hellinger distance is that it doesn't require the distributions $P$ and $Q$ to have the same support. The support of a distribution is defined as the subset of the distribution function domain with non-vanishing probability values. This property ensures that reasonable distance values can be obtained even when the probability mass of one distribution is concentrated around a few data points. This scenario is relevant to GHZ states where the noiseless distribution assumes equal probability values for only two bitstring outcomes, while the noisy GHZ histogram attains a more smeared-out distribution for all states. The Hellinger distance effectively accounts for outcomes with noiseless zero-probability contributions, a feature that the classical fidelity neglects. Thus, the Hellinger distance will be used in the following to evaluate the quality of our simulation results compared to the real measured ones.

\subsection{Benchmark Circuits}
\label{subsec:benchmark circuits}
With our model, we aim to accurately describe the execution of noisy circuits of different depths. Therefore, we chose a variety of benchmark circuits to evaluate the accuracy of our noise model, including small, medium, and large circuits. Moreover, we used both random and structured circuits to avoid any bias due to the simplification of the noise occurring in random circuits \cite{wallman2016noise} and to show that our model describes structured circuits as well. The circuits already match the target gate set of the associated hardware (see. Sec.~\ref{sec:iqm_qpu}), i.e. contain only $\texttt{PRX}$ and $\texttt{CZ}$ for the IQM QPU, and $\texttt{U1}$, $\texttt{U2}$, $\texttt{U3}$ and $\texttt{CNOT}$ for IBM Q Melbourne.

\begin{figure*}
\includegraphics[width=.99\textwidth]{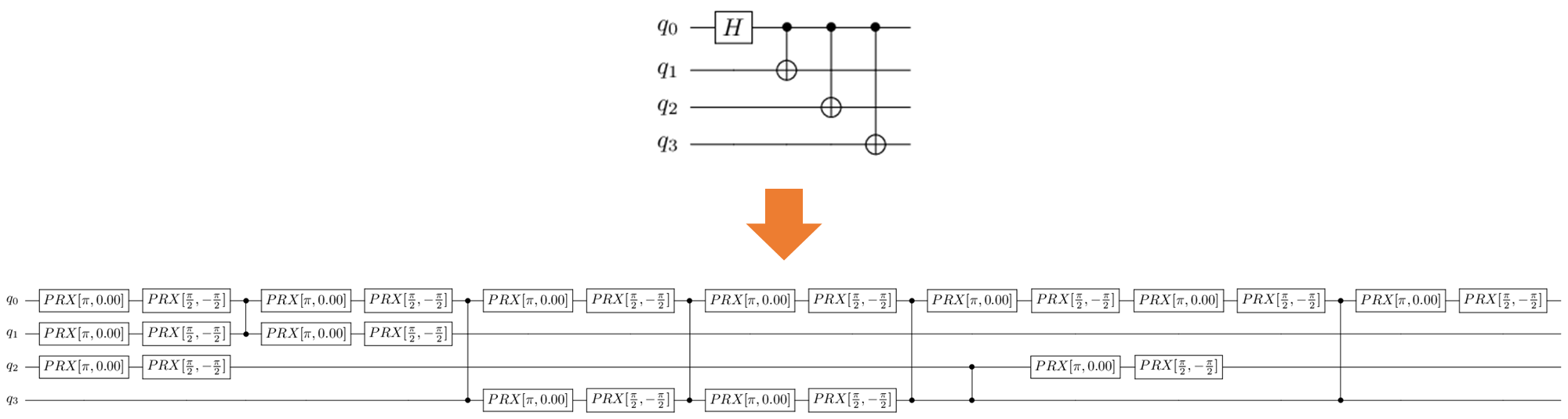}
\caption{Compilation of an input GHZ-4 circuit (top) into an IQM-compatible circuit (bottom). The circuit was translated using a compiler designed for the IQM-20-qubit QPU.}
\label{fig:compilation}
\end{figure*}

To generate these benchmark circuits, a compiler was employed to transform the initial logical circuits into circuits with an adapted topology and gateset. An example is shown in Fig.~\ref{fig:compilation}. As illustrated, the GHZ-4 circuit, comprising of one Hadamard gate $\texttt{H}$ and three $\texttt{CNOT}$ gates was transformed. As a first step, the circuit's topology has been adapted to the QPU by using a SWAP-insertion ansatz~\cite{swap}, which has been implemented using Eviden's Qaptiva framework and is referred to as Nnizer~\cite{nnizer}. As a final step, the compiler translates all non-native gates, such as the \texttt{H} and \texttt{CNOT} gate, to native gates of the QPU, i.e. $\texttt{PRX}$ and $\texttt{CZ}$ gates.

In particular, the benchmark circuits themselves can be divided into four classes: Greenberger-Horne-Zeilinger (GHZ) circuits~\cite{Greenberger2009} for two to seven qubits, a random unitary (RU) circuit and one QAOA circuit for the IQM device, and quantum walk circuits for four to fifteen qubits~\cite{GeorgopoulosUNM} for IBM Q Melbourne. For GHZ circuits starting at eight qubits, we observed that the quality of the measurements began to decrease, and comparing them to our noise model would no longer make sense. The main circuit resources per benchmark circuit are listed in Table~\ref{tab:circuit_parameters}. The circuit representations for each benchmark circuit are linked in App.~\ref{app1}.

\begin{table}[]
    \vspace{0.5cm}
    \centering
    \begin{tabular}{c|c|c|c|c} \hline
        Circuit & Qubits & Depth & 1QB Gates & 2QB Gates \\  \hline
        GHZ-2 & 2 & 5 & 6 & 1\\
        GHZ-3 & 4 & 16 & 22 & 5 \\
        GHZ-4 & 4 & 19 & 26 & 6 \\
        GHZ-5 & 5 & 20 & 30 & 7 \\
        GHZ-6 & 6 & 36 & 46 & 11 \\
        GHZ-7 & 7 & 46 & 62 & 15 \\
        RU & 5 & 33 & 48 & 13 \\
        QAOA  & 5 & 87 & 110 & 37 \\
        QW-2~\cite{GeorgopoulosUNM} & 4 & 44 & 19 & 35 \\
        QW-3~\cite{GeorgopoulosUNM} & 6 & 194 & 83 & 178 \\
        QW-4~\cite{GeorgopoulosUNM} & 11 & 409 & 175 & 423 \\
        QW-5~\cite{GeorgopoulosUNM} & 14 & 733 & 297 & 811 \\
        QW-6~\cite{GeorgopoulosUNM} & 15 & 1062 & 449 & 1137 \\
        \hline
    \end{tabular}
    \caption{The number of qubits, circuit depth, and the number of single-and two-qubit gates for the benchmark circuits investigated during this work.}
    \label{tab:circuit_parameters}
\end{table}

\subsection{Results}
\subsubsection{GHZ, random unitary and QAOA circuits}
%\mariuscom{TODO: add comment to transpiler in main text}

In this section, we will compare the simulation results of our noise model with the ones obtained from the real IQM quantum computing hardware as presented in Sec.~\ref{sec:iqm_qpu}. For the purpose of our comparison, we will be using the circuits described in Sec.~\ref{subsec:benchmark circuits} and the Hellinger distance as a quantitative measure of agreement (see Sec.~\ref{subsec:metric}).
The real measurement results were obtained by running each circuit on the IQM hardware for 10,000 times to obtain the respective probability distribution, i.e., the probabilities associated with the final quantum states.

The simulation results were obtained by implementing the noise model described in Sec.~\ref{sec:noise_model} using the Qaptiva software framework and subsequently simulating it on emulators included in Qaptiva. 

To ensure a realistic comparison, a deterministic simulation yielding a probability distribution for every single state has been used. To get an exact comparison with the hardware results, we then multiplied the probability of every single state by 10,000. The histograms illustrating the outcomes of both the simulations and the measurements are presented in Fig.~\ref{fig:results_others} for two GHZ circuits, one random unitary and one QAOA circuit. A comprehensive summary of the respective Hellinger distances is provided in Table~\ref{tab:Hellinger results}.

Both the histograms and the Hellinger distances show good agreement between our noise model and the actual experimental data. Especially for the smaller circuits -- namely GHZ-2 to GHZ-5 -- the agreement of the results is almost perfect with Hellinger distances below $0.1$. For GHZ-6 and GHZ-7 the noise model still accurately reflects the real hardware, with the worst Hellinger distance of $0.115$ for the GHZ-6 circuit. The random unitary and QAOA circuit again show Hellinger distances well below $0.1$, highlighting that even for larger circuit depths, our noise model matches very well with the measurements performed on the real hardware. Note that for the QAOA circuits, the hardware outputs are close to random sampling due to the large depth of the circuits. However, our simulations nicely reflect this behavior, proving that the noise model is capable of simulating very large and noisy circuits and giving good evaluations of the result quality of the real hardware.

We also observe discrepancies between the simulated and measured data. An example is the asymmetry of the target state for the GHZ-3 results (Fig.~\ref{fig:results_others} top right), which is not resolved as strongly by the simulation as it is in the real hardware. For the QAOA circuit (Fig.~\ref{fig:results_others} bottom right) the probability distribution is biased towards lower excitation states in the measurement, which is not reproduced as strongly in the simulation. These deviations between simulation and real data are most likely due to more complex noise sources. The most reasonable candidates for superconducting hardware are coherent errors (e.g. overrotations), leakage out of the computational subspace, and crosstalk \cite{papivc2023fast}.

However, an analysis of the results of~\cite{Georgopoulos2021} reveals that our noise model outperforms competing models when comparing different noise models. This is evident in all circuit sizes, particularly when utilizing calibrated hardware parameters instead of fitted ones. These findings demonstrate the efficacy of our model in accurately simulating real quantum computing hardware for circuits of varying depths. These results will be presented in the next subsection.

\begin{table}[]
    \centering
    \begin{tabular}{c|c} \hline
         Circuit & Hell. dist. \\ \hline
         GHZ-2 & $0.040$ \\
         GHZ-3 & $0.044$ \\ 
         GHZ-4 & $0.055$ \\
         GHZ-5 & $0.073$ \\
         GHZ-6 & $0.115$ \\
         GHZ-7 & $0.104$ \\
         RU & $0.055$ \\
         QAOA & $0.071$ \\
         %QAOA (max grad.) & $0.078$ \\
         %QAOA (final) & $0.112$ \\
         \hline
    \end{tabular}
    \caption{Hellinger distances for the benchmark circuits performed on the IQM 20-qubit QPU and averaged over a total of 50 measurements.}
    \label{tab:Hellinger results}
\end{table}

\subsubsection{Comparison to other noise models}
To gain more insight into whether our proposed noise model performs well, we also made a comparison with another approach, namely the so-called unified noise model (UNM)~\cite{Georgopoulos2021}. We studied the same circuits and implemented an emulated version of IBM Q Melbourne, with a native gateset consisting of \texttt{U1}, \texttt{U2}, \texttt{U3} as the single-qubit gates and \texttt{CNOT} as the native two-qubit gate. The noise parameters for this QPU are listed in Table~\ref{tab:noise_parameters}. This way, we are able to compare our noise model with the results obtained with the UNM. In addition, we take the Qiskit composite model, which has also been studied in~\cite{Georgopoulos2021}, as a further noise model for our benchmark analysis.

\begin{figure}[!t]
\includegraphics[width=.45\textwidth]{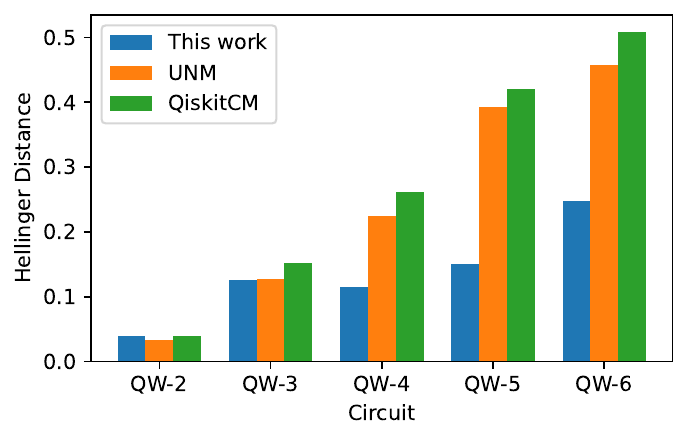}
\caption{Hellinger distances for the quantum walk circuits from this work and~\cite{Georgopoulos2021} that have been measured on IBM Q Melbourne. We compare our approach with the unified noise model (UNM) and the Qiskit composite model (QiskitCM). The Hellinger distances for the latter two models were taken from~\cite{Georgopoulos2021}.}
\label{fig:QWResults}
\end{figure}

Fig.~\ref{fig:QWResults} shows the results for the Hellinger distances for the quantum walk circuits. Comparing these values with the noise models presented in~\cite{Georgopoulos2021} -- such as the UNM and the Qiskit composite model -- we can see that the two only circuits for which our proposed approach performs similarly is the 2- and 3-qubit quantum walk (QW-2 and QW-3) when compared to the UNM and the Qiskit composite model, whereas for all other cases (4- to 6-qubit quantum walks, i.e. QW-4, QW-5 and QW-6), we can see a good improvement of about 50\%. An explanation for this observation could be that, especially for larger circuits, setting the relaxation and dephasing channel to simulate idle noise only on idle qubits might be a better approach.

In Fig.~\ref{fig:qw_results} of App.~\ref{app3}, the counts of the measurements performed on IBM Q Melbourne and the counts obtained from the noise model for the quantum walk circuits are presented. Specifically, for the 2-, 3-, and 4-qubit quantum walk circuits (QW-2, QW-3, and QW-4), there is a substantial agreement between the measurement and simulation results. However, for the more complex circuits -- namely QW-5 and QW-6 -- with depths exceeding 700 and 1000, respectively, the Hellinger distance increases. This is completely expected as any discrepancy between the model and experiment is amplified for the longer circuits. As illustrated in the lower graph depicting the QW-6 circuit in Fig.~\ref{fig:qw_results} of App.~\ref{app3}, there is a substantial count for the all-zero state, while the simulation indicates a significantly lower count. Despite this discrepancy in the all-zero state, simulation and measurement outcomes are in good agreement for all other states.

\begin{table}[]
    \vspace{0.5cm}
    \centering
    \begin{tabular}{c|c} \hline
        Parameter & Avg. value \\  \hline
        $\mathcal{F}_\texttt{1QB}$ & \SI{99.99}{\%}\\
        $\mathcal{F}_\texttt{2QB}$ & \SI{96.83}{\%} \\
        $T_\texttt{1QB}$ & \SI{100}{ns}\\
        $T_\texttt{2QB}$ & \SI{500}{ns}\\
        $T_1$ & \SI{56.15}{\mu s} \\
        $T_2$ & \SI{56.01}{\mu s} \\
        $\epsilon_{\rm meas}^{0}$ & \SI{7.61}{\%}\\
        $\epsilon_{\rm meas}^{1}$ & \SI{7.61}{\%}\\ 
        \hline
    \end{tabular}
    \caption{Average noise parameters for IBM Q Melbourne. Here, \texttt{1QB} denotes the single-qubit gates \texttt{U1}, \texttt{U2} and \texttt{U3}, where \texttt{2QB} denotes the two-qubit gate \texttt{CNOT}. The single- and two-qubit gate times $T_\texttt{1QB}$ and $T_\texttt{2QB}$ have been taken from~\cite{Martinez2021}. All other noise parameters have been taken from~\cite{Georgopoulos2021}.}
    \label{tab:noise_parameters}
\end{table}

\iffalse
\begin{table}[]
    \centering
    \begin{tabular}{c|c|c|c} \hline
         Circuit & This work & UNM & QiskitCM \\ \hline
         QW-2 & 0.039 & 0.033 & 0.040 \\
         QW-3 & 0.126 & 0.127 & 0.152 \\ 
         QW-4 & 0.115 & 0.224 & 0.262 \\
         QW-5 & 0.151 & 0.393 & 0.421 \\
         QW-6 & 0.247 & 0.457 & 0.509 \\
         \hline
    \end{tabular}
    \caption{Hellinger distances for the quantum walk circuits from~\cite{Georgopoulos2021} that have been measured on IBM Q Melbourne. We compare our approach with the unified noise model (UNM) and the Qiskit composite model (QiskitCM). The Hellinger distances for the latter two models were taken from~\cite{Georgopoulos2021}.}
    \label{tab:QWResults}
\end{table}
\fi

\section{Conclusion}
\label{sec:conclusion}

In this work, we present a noise model that can describe many common gate-based quantum computing hardware platforms by using the typical noise parameters supplied by quantum hardware vendors. 
% One substantial difference compared to previous work is our modeling of idling noise. These channels are only applied to qubits that are actually in the idle state, i.e. when no gates are acting on these qubits. 
We evaluated the accuracy of our model by comparing the simulation results of our noise model with the measurement results of a 20-qubit superconducting quantum computing hardware. Furthermore, we compared our noise model to the one developed by Georgopoulos et. al.~\cite{Georgopoulos2021} and demonstrated that, particularly for large circuits with depths greater than 400, we could reduce the Hellinger distances by about 50\%. In addition, we have shown that the simulations predict the behavior of real hardware in a truthful way for different types of circuits and depths. We attribute this increase in performance to a more accurate modeling of the noise on idling qubits.

Subsequent studies should explore the incorporation of more advanced error channels that yield minor discrepancies between simulation and real results. The noise model can be used for the evaluation and optimization of quantum algorithms prior to their execution on actual quantum computing hardware. Additionally, it can serve as a robust noise model for applications such as error mitigation. %While preparing this paper, Bravo-Montes et al.~\cite{bravo2024methodology} published a paper demonstrating that extracting topology and calibration parameters from real quantum processors allows one to select the ideal combination of noise channels that best reproduces the device's behavior.
%It has been demonstrated that predictions of a satisfactory degree of fidelity can be obtained by employing a distinct hardware platform and utilizing the model described in~\cite{bravo2024methodology}.

In the near future, one could try to improve the presented noise model further by including for instance crosstalk and leakage at the gate level. Additionally, a study of this noise model on different architectures, such as trapped-ion or neutral atom quantum computers should be performed.

\begin{acknowledgments}
This work was supported by the Q-Exa project funded by the German Federal Ministry of Research, Technology and Space (grant No.~13N16064). The authors thank the IQM team for regularly providing the noise data and Corentin Bertrand for engaging in helpful discussions throughout this work.
\end{acknowledgments}

\begin{appendix}
\section{Noise Data for the IQM 20-Qubit Chip}
\label{app2}

Table~\ref{tab:noise_parameters_1qb} and Table~\ref{tab:noise_parameters_2qb} give more information about the noise parameters on each qubit and each qubit pair.

\begin{table}[]
    \vspace{0.5cm}
    \centering
    \begin{tabular}{c|c|c|c|c|c} \hline
        Qubit No. & $\mathcal{F}$ (\%) & $T_1$ ($\mu$s) & $T_2$ ($\mu$s) & $\epsilon_{\rm meas}^{0}$ (\%) & $\epsilon_{\rm meas}^{1}$ (\%) \\  \hline
        0 &  99.51 & 39.3 & 1.8 & 3.10 & 9.85 \\
        1 &  99.94 & 63.4 & 3.1 & 2.45 & 1.90 \\
        2 &  99.89 & 42.7 & 4.8 & 2.65 & 3.95 \\
        3 &  99.85 & 43.5 & 2.9 & 3.65 & 5.90 \\
        4 &  99.89 & 46.0 & 2.4 & 2.05 & 3.40 \\
        5 &  99.91 & 49.3 & 2.5 & 2.00 & 2.65 \\
        6 &  99.62 & 55.4 & 1.9 & 2.60 & 3.50 \\
        7 &  99.90 & 36.1 & 3.0 & 1.95 & 2.60 \\
        8 &  99.88 & 33.1 & 2.1 & 2.20 & 4.55 \\
        9 &  99.90 & 47.0 & 5.1 & 3.70 & 3.70 \\
        10 & 99.87 & 34.8 & 2.6 & 1.85 & 6.95 \\
        11 & 99.91 & 64.9 & 4.7 & 4.55 & 7.15 \\
        12 & 99.89 & 23.1 & 5.5 & 1.20 & 3.30 \\
        13 & 99.92 & 45.8 & 2.1 & 1.50 & 1.85 \\
        14 & 99.62 & 7.0  & 1.8 & 3.50 & 7.80 \\
        15 & 99.91 & 46.3 & 5.0 & 3.35 & 4.00 \\
        16 & 99.88 & 46.8 & 2.7 & 4.05 & 18.75 \\
        17 & 99.88 & 35.1 & 3.1 & 2.15 & 3.55 \\
        18 & 99.92 & 39.5 & 4.9 & 2.30 & 3.40 \\
        19 & 99.84 & 36.4 & 1.9 & 2.40 & 2.95 \\
        \hline
    \end{tabular}
    \caption{Single-qubit gate parameters (fidelity $\mathcal{F}$, relaxation time $T_1$, depolarization time $T_2$ and measurement errors for state 0 and 1 $\epsilon_{\rm meas}^{0/1}$) for all qubits of the IQM 20-qubit chip. The single-qubit gate time is set to $T_\texttt{PRX}=\SI{20}{ns}$ for all qubits.}
    \label{tab:noise_parameters_1qb}
\end{table}

\begin{table}[]
    \vspace{0.5cm}
    \centering
    \begin{tabular}{c|c} \hline
        Qubit Pair & $\mathcal{F}$ (\%) \\  \hline
        0, 1   & 99.29 \\
        0, 3   & 99.02 \\
        2, 3   & 95.87 \\
        2, 7   & 99.13 \\
        1, 4   & 99.31 \\
        3, 4   & 97.32 \\
        3, 8   & 98.80 \\
        7, 8   & 99.22 \\
        7, 12  & 99.10 \\
        4, 5   & 99.38 \\
        4, 9   & 99.23 \\ 
        8, 9   & 99.38 \\
        8, 13  & 99.12 \\
        12, 13 & 99.28 \\
        5, 6   & 99.47 \\
        5, 10  & 99.13 \\
        9, 10  & 99.25 \\
        9, 14  & 98.71 \\ 
        13, 14 & 98.83 \\
        13, 17 & 99.10 \\
        6, 11  & 97.69 \\
        10, 11 & 92.28 \\
        10, 15 & 98.81 \\
        14, 15 & 98.54 \\
        14, 18 & 98.54 \\
        17, 18 & 99.02 \\
        11, 16 & 97.91 \\
        15, 16 & 99.29 \\
        15, 19 & 98.87 \\
        18, 19 & 98.97 \\
        \hline
    \end{tabular}
    \caption{Two-qubit gate parameters (fidelity $\mathcal{F}$) for all qubits pairs of the IQM 20-qubit chip. The two-qubit gate time is set to $T_\texttt{CZ}=\SI{40}{ns}$ for all qubits.}
    \label{tab:noise_parameters_2qb}
\end{table}

\section{Benchmark Circuits}
\label{app1}
As mentioned above, we investigated the behavior of our noise model by benchmarking different types of circuits, namely a set of GHZ circuits, a random unitary circuit and a QAOA circuit. These circuits in QASM format can be found on https://zenodo.org/records/16082428.

\iffalse
\begin{figure*}
\includegraphics[width=.39\textwidth]{GHZ_2.png}
\vspace{20px}
\includegraphics[width=.79\textwidth]{GHZ_3.png}
\vspace{20px}
\includegraphics[width=.79\textwidth]{GHZ_4.png}
\vspace{20px}
\includegraphics[width=.89\textwidth]{GHZ_5.png}
\vspace{20px}
\includegraphics[width=.99\textwidth]{GHZ_6.png}
\vspace{20px}
\includegraphics[width=.99\textwidth]{GHZ_7.png}
\caption{\textit{GHZ circuits that have been investigated during this work. From top to bottom: GHZ-2, GHZ-3, GHZ-4, GHZ-5, GHZ-6, and GHZ-7.}}
\label{fig:benchmark_ghz}
\end{figure*}

\begin{figure*}
\includegraphics[width=.99\textwidth]{RUN.png}
\caption{\textit{Random unitary circuit that has been investigated during this work.}}
\label{fig:benchmark_run}
\end{figure*}

\begin{figure*}
\includegraphics[width=.99\textwidth]{QAOA.png}
\caption{\textit{The QAOA circuit that has been investigated during this work. }}
\label{fig:benchmark_run}
\end{figure*}

\begin{figure*}
\includegraphics[width=.99\textwidth]{QW-2.png}
\caption{\textit{A 2-qubit quantum walk circuit that has been investigated during this work.}}
\label{fig:benchmark_run}
\end{figure*}
\fi

\section{Results for different Benchmark Circuits}
\label{app3}
\begin{figure*}[!ht]
\includegraphics[width=.49\textwidth]{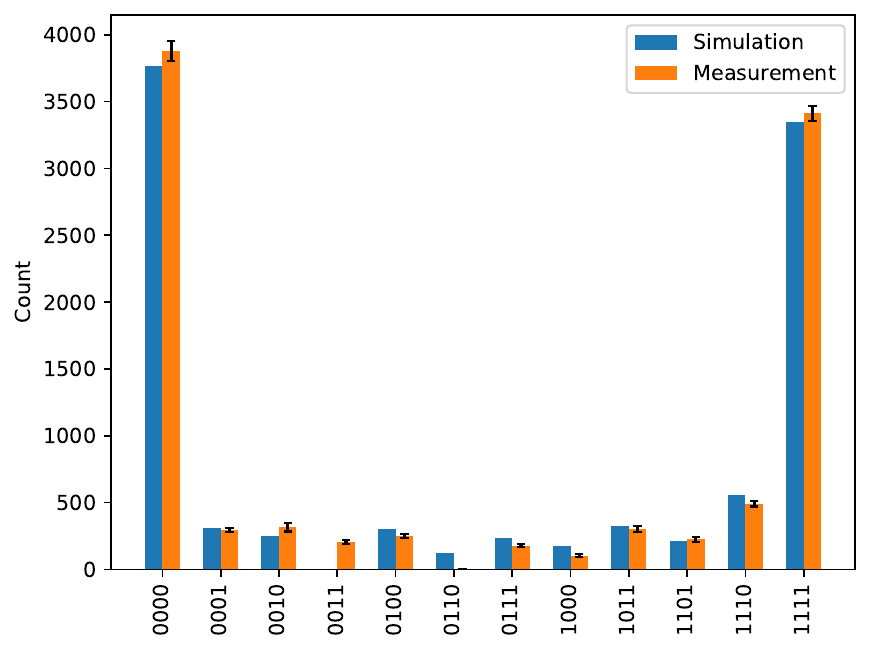}
\includegraphics[width=.49\textwidth]{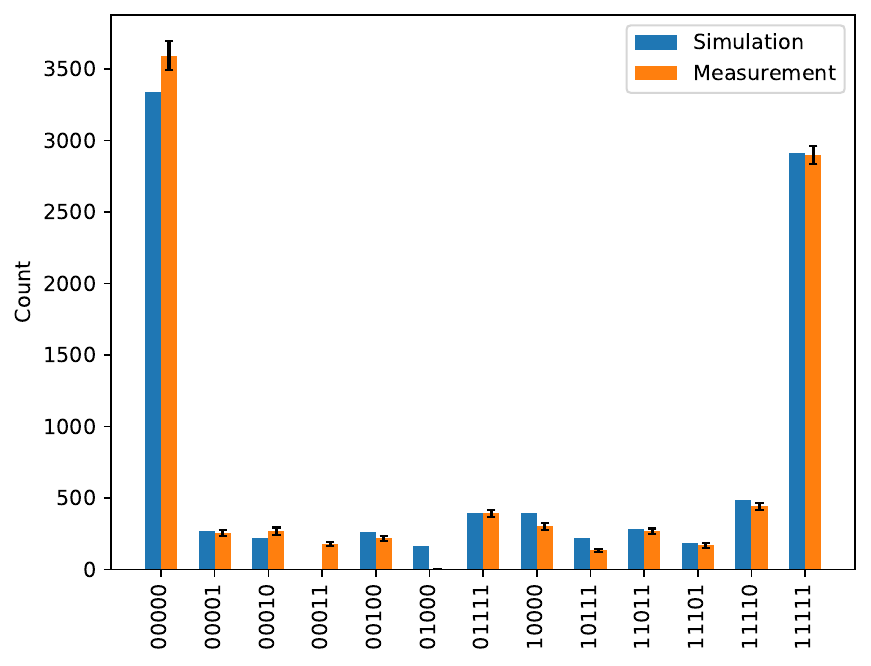}
\includegraphics[width=.49\textwidth]{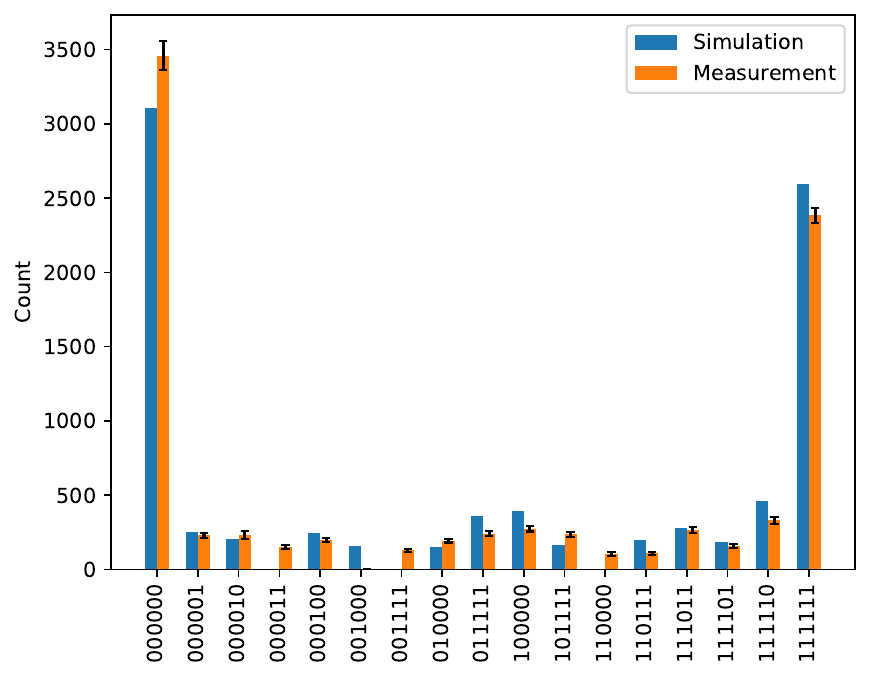}
\includegraphics[width=.49\textwidth]{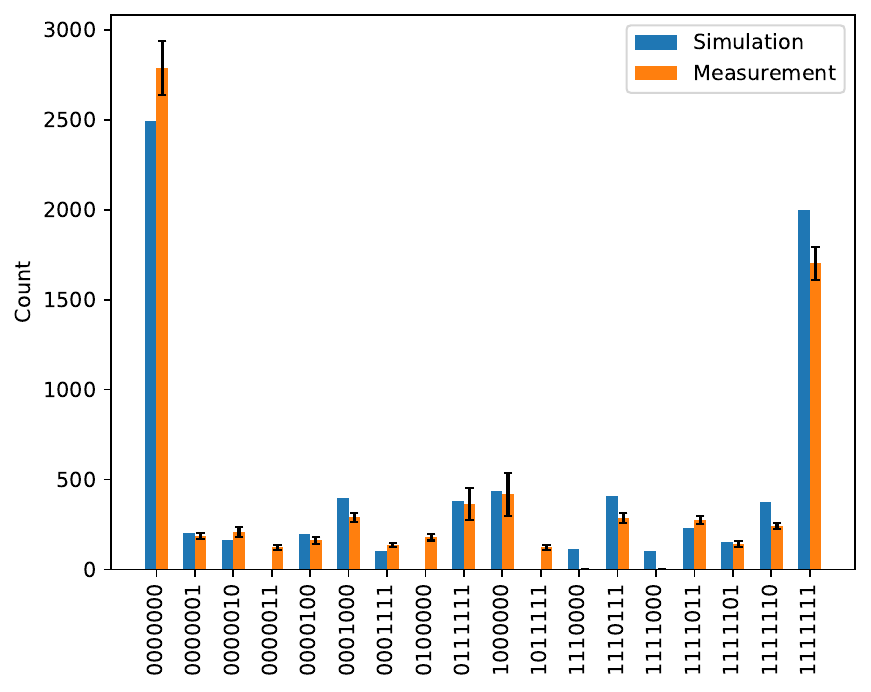}
\caption{Comparison of the results of our noise model simulations and the measurement result obtained on the IQM 20-qubit QPU for the GHZ circuits: GHZ-4 (top left), GHZ-5 (top right), GHZ-6 (bottom left), and GHZ-7 (bottom right). To get the simulated results, a deterministic simulation was performed which generates a probability distribution for all states. The state probabilities have then been multiplied by 10,000 to get the counts.
The results on real hardware were obtained by repeating 10,000 circuit executions (shots) 50 times. The blue bars show the counts obtained from the simulation for the respective states, and the orange bars show the counts measured in the experiment. The error bars represent the standard deviations occurring from the repetitions of the measurements. States with counts below 100 have been neglected for visualization reasons.}
\label{fig:results_GHZ_large}
\end{figure*}

\begin{figure*}[!ht]
\includegraphics[width=.49\textwidth]{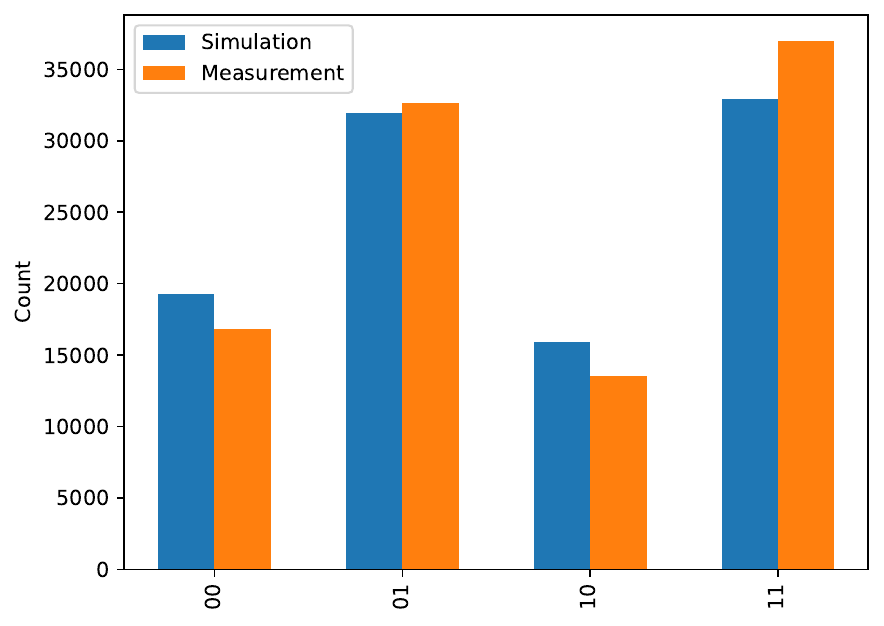}
\includegraphics[width=.49\textwidth]{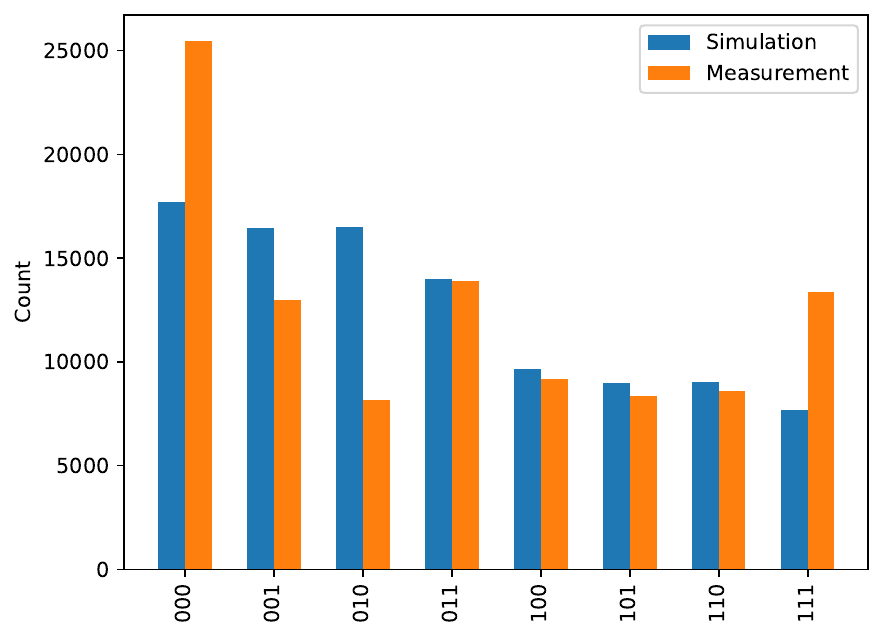}
\includegraphics[width=.49\textwidth]{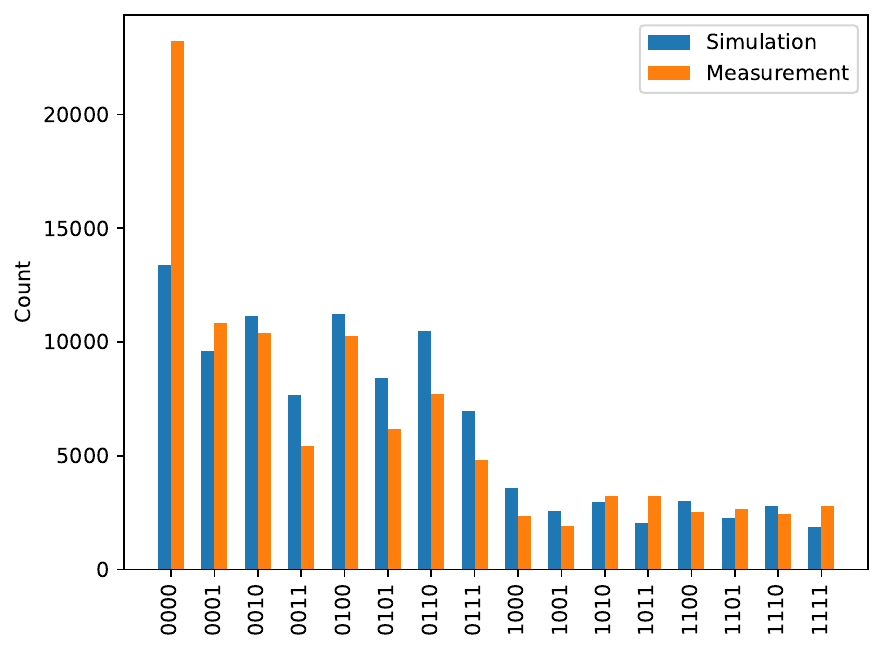}
\includegraphics[width=.49\textwidth]{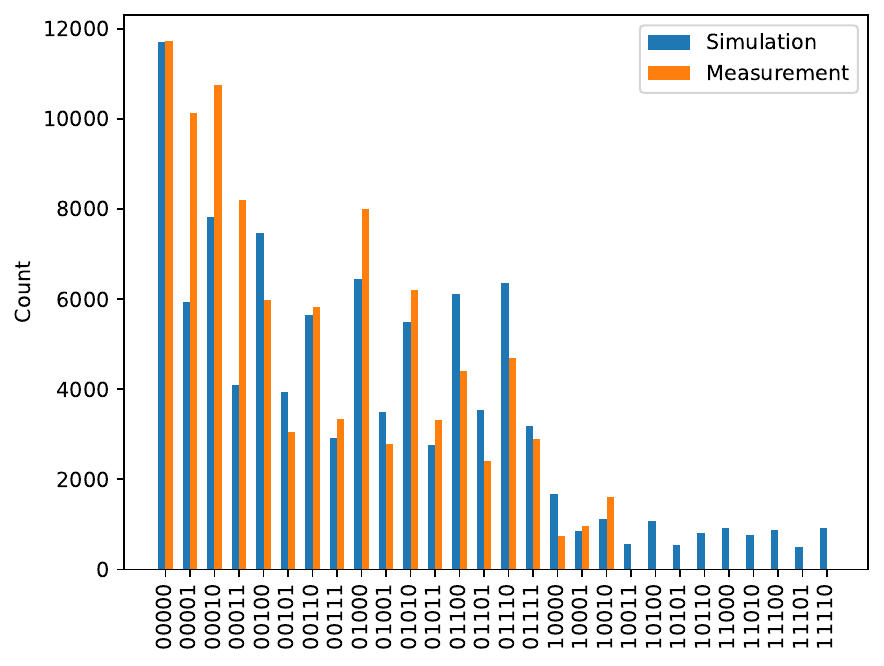}
\includegraphics[width=.49\textwidth]{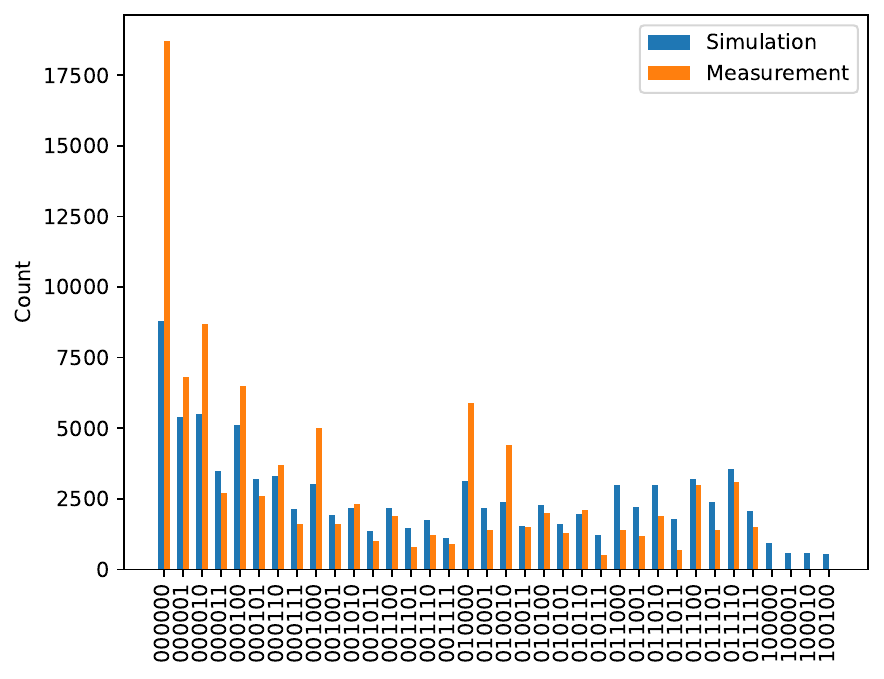}
\caption{Comparison of the results obtained from our noise model simulations and the measurement result obtained from IBM Q Melbourne for quantum walk circuits: QW-2 (top left), QW-3 (top right), QW-4 (center left), QW-5 (center right), and QW-6 (bottom). Both simulation and measurement results were obtained using 100,000 circuit executions (shots). The blue bars show the counts obtained from the simulation for the respective states and the orange bars the counts obtained from the real QPU. In this case, the uncertainties of the experimental results are in the order of approximately 100 shots and are therefore not visible in the plots. States with counts below 500 have been neglected for visualization reasons. The measurement data was taken from~\cite{Georgopoulos2021}.}
\label{fig:qw_results}
\end{figure*}

In this section, we present the results obtained from the simulated and real QPUs. Fig.~\ref{fig:results_GHZ_large} shows the simulations of the circuits GHZ-4 to GHZ-7, while Fig.~\ref{fig:qw_results} highlights the results for the quantum walk circuits QW-2 to QW-6.

\end{appendix}

\clearpage
\bibliography{Bibliography}

\end{document}